\documentclass[aps,amsfonts,pra,showpacs,nobibnotes,nofootinbib,preprint]{revtex4}
\usepackage{graphicx}
\begin{document}
\newcommand{\beq}{\begin{equation}}
\newcommand{\eeq}{\end{equation}}
\newcommand{\barr}{\begin{eqnarray}}
\newcommand{\earr}{\end{eqnarray}}
\newcommand{\tavg}[1]{\left< #1 \right>_T}
\def\figwidth{7.5cm}
\newcommand{\Mfunction}[1]{#1}
\newcommand{\asfigure}[3]{\includegraphics[width=#3]{#1}}
\def\cH{{\cal H}}
\def\cT{{\cal T}}
\def\cP{{\cal P}}
\def\cD{{\cal D}}
\def\cG{{\cal G}}
\def\cV{{\cal V}}
\def\cF{{\cal F}}
\def\cL{{\cal L}}
\def\cR{{\cal R}}
\def\cU{{\cal U}}
\def\cS{{\cal S}}
\def\cO{{\cal O}}
\def\cE{{\cal E}}
\def\bfA{{\bf A}}
\def\bfG{{\bf G}}
\def\bfn{{\bf n}}
\def\bfr{{\bf r}}
\def\bfV{{\bf V}}
\def\bft{{\bf t}}
\def\bfM{{\bf M}}
\def\bfS{{\bf S}}
\def\bfP{{\bf P}}
\def\vecj{\vec{j}}
\def\vnabla{\vec{\nabla}}
\def\vS{\vec{S}}
\def\vn{\vec{n}}
\def\vf{\vec{f}}
\def\vecr{\vec{r}}
\def\svn{{\vec{n}}}
\def\bra#1{\langle #1 |}
\def\ket#1{| #1 \rangle}
\newcommand{\p}{\partial}
\def\coltwovector#1#2{\left({#1\atop#2}\right)}
\def\upp{\coltwovector10}
\def\downn{\coltwovector01}
\def\Ord#1{{\cal O}\left( #1\right)}
\def\bmp{\mbox{\boldmath $p$}}
\def\rhobar{\bar{\rho}}
\renewcommand{\Re}{{\rm Re}}
\renewcommand{\Im}{{\rm Im}}
\renewcommand{\theequation}{\thesection.\arabic{equation}}
\def\ask{\marginpar{?? ask:  \hfill}}
\def\fin{\marginpar{fill in ... \hfill}}
\def\note{\marginpar{note \hfill}}
\def\check{\marginpar{check \hfill}}
\def\discuss{\marginpar{discuss \hfill}}
\title{Casimir Effects: an Optical
Approach
\\
II.  Local Observables and Thermal Corrections
}\author{A.~Scardicchio}\email{scardicc@mit.edu}
\author{R.~L.~Jaffe}\email{jaffe@mit.edu}
\affiliation{Center for Theoretical Physics, \\ Laboratory for Nuclear Sciences and Physics Department\\
Massachusetts Institute
  of Technology \\ Cambridge, MA 02139, USA}
\begin{abstract}
    \noindent We recently proposed a new approach to the Casimir effect based on
    classical ray optics (the ``optical approximation"). In this paper we show
    how to use it to calculate the local observables of the field theory.
    In particular we study the energy-momentum tensor and
    the Casimir pressure. We work three examples in detail:
    parallel plates, the Casimir pendulum and a sphere opposite a
    plate. We also show how to calculate thermal corrections,
    proving that the high temperature `classical limit' is indeed valid for any smooth geometry.
\end{abstract}
\vspace*{-\bigskipamount} \preprint{MIT-CTP-3652} \maketitle
\setcounter{equation}{0}
\section{Introduction}

The Casimir effect\cite{Casimir, Lifshitz, MT, MPnew,expt1} is a
manifestation of the quantum fluctuations of a quantum field at a
macroscopic level. Experiments on Casimir forces are precise tests
of one of the less intuitive predictions of field theory. For a
theoretician, predicting the outcome of these experiments is a
worthy challenge. Hence it seems somewhat astonishing that an
exact solution exists only for infinite, parallel plates
case\cite{Casimir}. Other formal solutions for geometries not made
of distinct rigid bodies free to move (like the wedge, the
interior of a sphere or of a rectangular box \cite{MT}) are
irrelevant for an experimental setup. Moreover in many such
solutions divergences have been discarded in a way that leaves the
result unrelated to practical materials and configurations
\cite{Graham:2002xq}. Interesting theoretical developments include
the method developed in \cite{Kardar} where the solution for
infinite periodic geometries is obtained as a series expansion in
the corrugation, and the numerical Montecarlo analysis in
\cite{Gies03}.

Amongst the various effects that an experimentalist must take in
account to interpret the data (e.g.\ finite conductivity,
temperature and roughness corrections) probably the most
challenging, interesting and full of connections with other
branches of physics and mathematics is the dependence of the force
on the geometry of the bodies. Calculating the Casimir force for
perfectly reflecting bodies in the end reduces to finding the
density of states (DOS) of the Scrh\"odinger Hamiltonian for the
equivalent billiard problem \emph{including the oscillatory ripple
on the averaged DOS}. This is an incredibly difficult problem in
spectral theory that still challenges mathematicians and
physicists today \cite{Uribe} and in essence is not solved beyond
the semiclassical approximation.

In this context we have introduced in Refs.\ \cite{pap1, opt1} a
method based on classical optics which has several virtues:
accuracy, uniform validity when a symmetry is born,
straightforward extension to higher spin fields, to non-zero
temperatures, to include finite reflectivity and, the main topic
of this paper, \emph{it can provide an approximation to local
observables.}

This paper is structured as follows: in Section
\ref{sec:enmomtens} we show how to cast the energy momentum tensor
into a sum over optical paths contributions and how to regulate
and analyze the divergences, ubiquitous in Casimir energy
calculations. Section III is dedicated to the analysis of the
three examples already studied in \cite{opt1} with pedagogical
intent. We study parallel plates, the Casimir torsion pendulum and
a sphere opposite a plate. In Section IV we show how to calculate
the same local observables and the free energy for a thermal state
and we prove (within the limits of our approximation) the
`classical limit' theorem \cite{MPnew,Feinberg}, which states that
at high $T$, Casimir forces become independent of $\hbar$ and
proportional to $T$. As far as we know this is the first time this
assertion can be generalized to geometries other than parallel
plates. We also study the example of parallel plates (finding the
known results) and of a sphere opposite a plate at non-zero
temperature.  We find evidence, again within the framework of the
optical approximation, that the low $T$ behavior of the Casimir
force is a difficult problem, qualitatively different from the
$T=0$ and high temperature cases.

\section{Local Observables}
\setcounter{equation}{0}
\label{sec:enmomtens}

Local properties of the quantum vacuum induced by the presence of
boundaries are of broad interest in quantum field theory
\cite{local}. For example gravity couples locally to the
energy-momentum tensor.  Vacuum polarization induces local charge
densities near boundaries, provided the symmetries of the theory
allow it.  Also, local densities are free from some of the cutoff
dependencies that plague many other Casimir effects.  Any local
observable that can be expressed in terms of the Greens function
can be estimated using the optical approach.  In this section we
study the energy, momentum and stress densities for a scalar
field.

Some local observables are not unambiguously defined
\cite{Weinberg}. For example the charge density (in a theory with
a conserved charge) is unambiguously defined while the energy
density, in general, is not (while its integral over the volume,
the total energy, is). In this paper we use the Noether definition
of the energy-momentum tensor, similar results would be obtained
with other interesting definitions.

\subsection{Energy-momentum tensor}

We study the Noether energy-momentum tensor of a free, real scalar
field $\phi$ in a domain $\cD$ with Dirichlet boundary conditions
(BC) on $\cS=\partial\cD$ made of (in general disconnected)
surfaces. Other BC\ (Neumann, Robin) can be discussed but for
simplicity we restrict ourselves to Dirichlet BC\ here.

The lagrangian is (we use $\hbar=c=1$)
\beq
\cL=\frac{1}{2}\p_\mu\phi\p^\mu\phi
    -\frac{1}{2}m^{2}\phi^{2},
\eeq
where Greek letters are used for 4-dimensional indices while the
vector notation will be used for spatial vectors.

The Noether energy-momentum tensor for this real scalar field is
\beq
T_{\mu\nu}=\frac{\partial \cL}{\partial(\partial^\mu
\phi)}\partial_\nu\phi-g_{\mu\nu}\cL
\eeq
\begin{equation}
    T_{\mu\nu}=
    \p_{\mu}\phi\ \p_{\nu}\phi-g_{\mu\nu}\frac{1}{2}\left(\p_{\alpha}\phi\ \p^\alpha\phi
    -m^{2}\phi^{2}\right)
    \label{tmunu}
\end{equation}
from which we identify the energy density $T_{00}$, the momentum
density $T_{0i}$, and the stress tensor $T_{ij}$. The definition
of these quadratic operators involves divergences that we will
regulate by point splitting. We hence replace quadratic operators
like $\phi(x)^2$ by $\lim_{x'\to x}\phi(x')\phi(x)$. The energy
density operator, for example, is
\begin{eqnarray}
T_{00}(x,t)&=&\lim_{x'\to
x}\left[\frac{1}{2}\partial_0\phi(x',t)\partial_0\phi(x,t)+\frac{1}{2}\vnabla'\cdot\vnabla\phi(x',t)\phi(x,t)+\frac{1}{2}m^2\phi(x',t)
\phi(x,t)\right]\nonumber\\
&=&\lim_{x'\to
x}\bigg[\frac{1}{2}\partial_0\phi(x',t)\partial_0\phi(x,t)-\frac{1}{2}\phi(x',t)\vnabla^2\phi(x,t)+\nonumber\\
&&+\frac{1}{2}m^2\phi(x',t)\phi(x,t)+
\frac{1}{2}(\vnabla'+\vnabla)\cdot\phi(x',t)\vnabla\phi(x,t)\bigg]
\label{eq:T00}
\end{eqnarray}
The field $\phi$ satisfies the free wave equation in $\cD$
\beq
\p^2\phi+m^2\phi=0
\eeq
and hence it can be decomposed into normal modes
\begin{equation}
\label{eq:decompo}
\phi(x,t)=\sum_j\frac{1}{\sqrt{2E_j}}\left(\psi_j(x)e^{-iE_jt}a_j+\psi_j^*(x)e^{iE_jt}a^{\dag}_j\right),
\end{equation}
where $\psi_j$ and $E_j$ are the eigenfunctions and eigenvalues of
the problem
\begin{eqnarray}
\label{eq:dirchletprobl}
(-\vnabla^2+m^2)\psi_j&=&E^2_j\psi_j\quad\mbox{for } x\in
\cD;\qquad \psi_j(x)=0 \quad \mbox{for }\; x\in\cS.
\end{eqnarray}
We also use the definition $E(k)=\sqrt{k^2+m^2}$, and
$E_j=\sqrt{k_j^2+m^2}$ so that the eigenvalue equation reads
\beq
-\vnabla^2\psi_j=k^2_j\psi_j,
\label{eq:dirchletproblk}
\eeq
and because of the positivity of the operator $-\vnabla^2$, the
spectrum $\{ E_j\}$ is contained in the half-line $\{E\geq m\}$.

We now introduce the propagator $G(x',x,k)$, defined as in Ref.\
\cite{opt1} to be the Green's function of the problem
(\ref{eq:dirchletprobl}) or (\ref{eq:dirchletproblk}):
\barr
(-\vnabla'^2-k^2)G(x',x,k)&=&\delta(x'-x)\nonumber\\
G(x',x)&=&0 \quad \mbox{for } x' \mbox{ or } x\in\cS,
\earr
which can be written using the spectral decomposition as
\begin{equation}
G(x',x,k)=\sum_n\frac{\psi_n(x')\psi_n(x)}{k^2_n-k^2-i\epsilon}
\end{equation}
In Ref.~\cite{pap1} we have developed an approximation for the
propagator $G(x',x,k)$ in terms of optical paths (closed, in the
limit $x'\to x$). The derivation can be found in Ref.~\cite{opt1},
the general result valid for $N$ spatial dimensions being
\begin{eqnarray}
    G_{\rm opt}(x',x,k)&=&\sum_r\frac{(-1)^{n_r}}{2^{N/2+1}\pi^{N/2-1}}\left(\ell_r\Delta_r\right)^{1/2}k^{N/2-1}H^{(1)}_{\frac{N}{2}-1}
    \left(k\ell_r\right),\nonumber \\
    \label{eq:unifsemicl}
    &\equiv&\sum_r G_r(x',x,k),
\end{eqnarray}
where $H$ is a Hankel function, $r$ labels the paths from $x$ to
$x'$, $n_r$ is the number of reflections of the path $r$,
$\ell_r(x',x)$ is its length and $\Delta_r(x',x)$ is the
\emph{enlargement factor} familiar from classical optics,
\begin{equation}
    \label{eq:deltar}
    \Delta_r(x',x)=\frac{d\Omega_{x}}{dA_{x'}}.
\end{equation}
$\Delta_r(x',x)$ is the ratio between the angular opening of a
pencil of rays at the point $x$ and the area spanned at the final
point $x'$ following the path $r$. For $N=3$ we have
\beq
\label{eq:Gopt3}
G_r(x',x,k)=(-1)^{n_r}\frac{\Delta^{1/2}_r(x',x)}{4\pi}e^{ik\ell_r(x',x)}.
\eeq

With this explicit form for the propagator $G$, we now have to
rewrite the elements of the quadratic operator $T_{\mu\nu}$ as
functions of $G$ and its derivatives. It is useful to pass from
the point-splitting to a frequency cutoff by inserting the latter
in the normal modes decomposition (\ref{eq:decompo}) as
\begin{equation}
e^{-k_j/\Lambda}=\int_0^\infty dke^{-k/\Lambda} 2k\;
\delta(k^2-k_j^2).
\end{equation}
The limit $x'\to x$ can then be exchanged with the $dk$ integral
and we get for the energy density,
\begin{equation}
\label{eq:T00opt}
\bra{0}T_{00}(x,t)\ket{0}=\int_0^\infty
dke^{-k/\Lambda}\frac{1}{2}E(k)\rho(x,k)+\int_0^\infty
dke^{-k/\Lambda} \frac{k}{2E(k)}\vnabla\cdot\vecj(x,k).
\end{equation}
The density $\rho$ and the vector $\vecj$ are defined as
\begin{eqnarray}
\label{eq:rhoxk}
\rho(x,k)&=&\frac{2k}{\pi}\Im\ G(x,x,k)\\
\vecj(x,k)&=&\lim_{x'\to x}\frac{1}{\pi}\Im \vnabla
G(x',x,k)=\frac{1}{2\pi}\Im\vnabla\ G(x,x,k).
\end{eqnarray}
$\cE$ is obtained by integrating $T_{00}$ over the whole volume
between the bodies:
\begin{equation}
\label{eq:volsurf}
\cE=\int_\cD d^3x\int_0^\infty dke^{-k/\Lambda}
\frac{1}{2}E(k)\rho(x,k)+\int_0^\infty
dke^{-k/\Lambda}\frac{k}{2E(k)}\int_\cS d\vS \cdot\vecj(x,k).
\end{equation}
We have turned the integral over the divergence of $\vecj$ into a
surface integral using Gauss's theorem. In the case of Dirichlet
or Neumann boundary conditions, since $d\vS\propto \vn$ we have
(here $\p_{\vn}\equiv \vn\cdot\vnabla$ and
$j_{\svn}=\vn\cdot\vecj$)
\begin{equation}
j_{\svn}(x,k)=\frac{1}{\pi}\Im\ \partial_{\svn} G(x,x,k)=0, \qquad
x\in \cS
\end{equation}
and the surface integral term disappears. It should be noted that
the vanishing of the $\vecj$ contribution to the total energy
relies on the continuity of the propagator for $x',x\in \cD$. In
some approximations, including the optical one, this continuity is
lost. Hence spurious surface terms arise on the boundary of
certain domains $\cD'\subset \cD$. This region is what in wave
optics is called the `penumbra' region. Diffractive contributions
are also not negligible in this region and they cancel the
discontinuities in $G$, hence eliminating the surface
terms.\footnote{As an example see Kirchoff's treatment of the
diffraction from a hole in Ref.~\cite{BornWolf}.} The surface
terms in the energy are hence of the same order of the diffractive
contributions which define the error in our approximation.

The divergence $\vnabla\cdot\vecj$ could also be eliminated from
$T_{00}$ by changing the energy-momentum tensor according to
\begin{equation}
\tilde{T}_{\mu\nu}=T_{\mu\nu}+\partial^\alpha\psi_{\alpha\mu\nu},
\end{equation}
with
\begin{equation}
\psi_{\alpha\mu\nu}=\frac{1}{2}\phi\left(g_{\mu\nu}\partial_\alpha-g_{\alpha\nu}\partial_\mu\right)\phi.
\end{equation}
The total energy $\cE$ and momentum are not affected by this
redefinition however the new tensor $T_{\mu\nu}$ is not symmetric.

It can be seen that the stress tensor $T_{ij}$ is normal on the
surface $\cS$ (for both Dirichlet and Neumann BC) so locally the
force on the surface is given by the pressure alone
\beq
\frac{d\vec{F}}{dS}=\vn P=\bra{0}T_{\vn,\vn}\ket{0}.
\eeq
The operator $T_{\svn\svn}$ regulated by point splitting is
\begin{eqnarray}
T_{\svn,\svn}(x,t)&=&\lim_{x'\to
x}\left[\partial'_{\svn}\phi'\partial_{\svn}\phi-\frac{1}{2}
g_{\svn,\svn}\left(\partial'_0\phi'\partial_0\phi-\vnabla'\phi'\cdot\vnabla\phi-m^2\phi^2\right)\right]\nonumber\\
&=&\lim_{x'\to
x}\Big[\partial'_{\vn}\phi'\partial_{\vn}\phi+\frac{1}{2}\left(\partial'_0\phi'\partial_0\phi+\phi'\vnabla^2\phi-m^2\phi^2\right)\nonumber\\
&-&\frac{1}{2} \left(\vnabla'+\vnabla\right)\phi'\vnabla\phi\Big]
\end{eqnarray}
where $\phi'$ is shorthand for $\phi(x',t)$. The second term in
brackets is zero when averaged over an eigenstate of the number
operator $\ket{\{n_j\}}$, by virtue of the equations of motion.
For Dirichlet BC\ the term $\phi\vnabla^2\phi=0$ on the
boundaries, so we have ($\vnabla=\vn\partial_{\svn}+\vnabla_t$)
\begin{equation}
\bra{0}T_{\svn,\svn}\ket{0}=\lim_{x'\to
x}\sum_j\frac{1}{4E_j}\left(\partial'_{\svn}\partial_{\svn}-\vnabla'_{t}\cdot\vnabla_{t}+k_j^2\right)\psi_{j}(x')\psi_j(x).
\end{equation}
Since also $\vnabla_t\psi_j(x)=0$ on the boundaries this
expression simplifies to
\begin{equation}
P(x)=\lim_{x'\to
x}\sum_j\frac{1}{4E_j}\partial'_{\svn}\partial_{\svn}\psi_j(x')\psi_j(x).
\end{equation}
This expression can be rewritten, in terms of the propagator $G$,
regulated by a frequency cutoff as we did for $T_{00}$,
\begin{equation}
P(x)=\lim_{x'\to x}\partial'_{\svn}\partial_{\svn}\int_0^\infty dk
e^{-k/\Lambda}\frac{k}{2\pi E(k)}\Im G(x',x,k).
\end{equation}
In this regulated expression we can exchange the derivatives,
limit and integral safely. Below we discuss what the divergences
are when $\Lambda\to\infty$ and how to interpret and dispose them.

All the above expressions are exact. Once the propagator $G$ is
known, we can calculate the energy-momentum tensor components from
them. However as discussed above in the interesting cases it is
difficult to find an exact expression for $G$ and some
approximations must be used.

For smooth impenetrable bodies we use the optical approximation to
the propagator developed in Ref.\ \cite{pap1,opt1} and recalled in
eq.\ (\ref{eq:unifsemicl}). This gives $G$ as a series of optical
paths and hence the pressure $P$ as a sum of optical paths
contributions
\barr
P&\simeq&\sum_r P_r\\
P_r&=&(-1)^{n_r}\lim_{x'\to
x}\partial'_{\svn}\partial_{\svn}\int_0^\infty dk
e^{-k/\Lambda}\frac{k}{2\pi
E(k)}\frac{\Delta_r^{1/2}(x',x)}{4\pi}\sin
\left(k\ell_r(x',x)\right),
\earr

An important feature of the optical approximation is that all
divergences are isolated in the low reflection terms whose
classical path length can vanish as $x',x\to\cS$. In practice only
the zeroth and first reflection are potentially divergent. Before
performing the integral in $k$ and taking $\Lambda\to\infty$ then
we have to put aside the divergent zero and one reflection terms
$P_0$ and $P_1$ for a moment (in the next section we will show how
their contributions are to be interpreted).

For the remaining families of paths (that we will denote as
$r\in\cR$) the integral over $k$ can be done and the limit
$\Lambda\to\infty$ taken safely. The result is finite and reads
\begin{equation}
P(x)= \sum_{r\in \cR}\lim_{x'\to
x}\partial'_{\svn}\partial_{\svn}(-1)^{n_r}\frac{\Delta_r^{1/2}(x',x)}{8\pi^2\ell_r(x',x)}.
\label{eq:pressurepaths}
\end{equation}
We can further simplify this expression. For simplicity let us
call $z$ the normal direction. Notice that for any sufficiently
smooth function $f(z',z)$ vanishing for either $z'$ or $z$ on the
surface $z=0$
\begin{equation}
\label{eq:trick1}
\partial_{z'}\partial_z
f(z',z){\Big|}_{z'=z=0}=\frac{1}{2}\partial_z^2f(z,z){\Big|}_{z=0}.
\end{equation}
The proof is trivial: consider that the lowest order term in the
expansion of $f(z',z)$ near $z',z=0$ is $\propto z'z$. The
propagator $G(x',x,k)$ satisfies all these properties and hence we
can use this result to get rid of the limit $x'\to x$ and assume
$x'=x$ from the beginning. We can therefore rewrite
Eq.~(\ref{eq:pressurepaths}) as,
\begin{equation}
\label{eq:press2}
P(x)= \sum_{r\in
\cR}(-1)^{n_r}\partial_{z}^2\frac{\Delta_r^{1/2}(x,x)}{16\pi^2\ell_r(x,x)}.
\end{equation}

Equation (\ref{eq:press2}) is one of the main results of this
paper. In Ref.\ \cite{opt1} we reduced the computation of Casimir
energy to a volume integral. The force is then found by taking the
derivative with respect to the distance between the bodies.
Calculating the pressure instead gives the force by means of just
a double integral of a local function. The problem is then
computationally lighter and sometimes (as we will see in the
examples) can even lead to analytic results.

Essentially the problem has been reduced to finding the lengths
and enlargement factors associated with the optical paths
\emph{for points close to the boundary}. In the case of the
pressure (eq.\ (\ref{eq:pressurepaths}) or (\ref{eq:press2})) it
is necessary to know their derivatives in the direction transverse
to the surfaces. We will see that this problem can be easily
tackled numerically when it cannot be solved analytically.

\subsection{Regulate and eliminate divergences}

As in the energy calculations \cite{opt1}, the only divergences
occurring in the pressure come from by paths whose lengths
$\ell\ll 1/\Lambda$, where $\Lambda$ is the plasma frequency of
the material. There are only two such families of paths: the zero
and one reflection paths. In this section we show that these
divergent contributions are independent of the distances between
the bodies. This fact is easily understood: in order for a path to
have arbitrarily small length all of its points must be on the
same body. So in order to study these terms we need only consider
a single, isolated body (and a massless field). We are also
careful in maintaining the double derivative $\partial^2_{z',z}$
since we are calculating the terms $P_0$ and $P_1$ separately.

For $r=0$, the zero reflection term, introducing an exponential
cutoff $\Lambda$ on the material reflection coefficient we obtain
\begin{equation}
\label{eq:pressure0}
P_{0}=\int_0^\infty e^{-k/\Lambda} dk \frac{k}{2\pi
E(k)}\lim_{x'\to x\in
\cS}\partial_{z'}\partial_{z}\left(\frac{\sin
k|z'-z|}{4\pi|z'-z|}\right)=\frac{\Lambda^4}{4\pi^2}.
\end{equation}
The same calculation for the $r=1$ or one reflection term gives:
\begin{equation}
\label{eq:pressure1}
P_{1}=\int_0^\infty e^{-k/\Lambda} dk \frac{k}{2\pi
E(k)}\lim_{x'\to
x\in\cS}\partial_{z'}\partial_{z}\left(-\frac{\sin
k|z'+z|}{4\pi|z'+z|}\right)=\frac{\Lambda^4}{4\pi^2}.
\end{equation}
Notice that these two terms are equal, so we could have
substituted $\partial_{z',z}\to\frac{1}{2}\partial^2_z$ for their
sum, after having properly regulated the divergence.

This positive, cut-off dependent pressure, $P_\Lambda\equiv
P_0+P_1$, must be dynamically balanced locally by a pressure
generated by the material, lest it collapse. Moreover the total
force obtained by integrating this quantity over the (closed)
surface $\cS$ of the whole body gives zero. However, if the space
around the body were inhomogeneous, as in the presence of a
gravitational field, a finite term survives the surface
integration, giving rise to a ``vacuum Archimedes effect'' in
which the pressure on one side is, due to gravitational effects,
larger than on the other side, so the body feels a net force. We
have analyzed this effect in detail in Ref.\ \cite{buoy} and
called it ``Casimir buoyancy''.

Finally note that another important element of this class of
quadratic operators is the Feynman propagator. In studying a field
theory in a cavity or in between impenetrable bodies (for example
hadrons as bags, photons in cavities or Bose-Einstein condensates
in traps), we can consider expanding the Feynman propagator in a
series of classical optical paths reflecting off the boundaries.
The first term, related to the direct path is the familiar free
propagator, the others give the finite volume corrections.

\section{Examples}
\setcounter{equation}{0}

In this section we calculate the Casimir force from the pressure,
using the formalism developed in the previous section, for three
examples that were already addressed in Ref.\ \cite{opt1} using
the energy method.

\subsection{Parallel Plates}
\label{sec:parplates}

The parallel plates calculation is a classic example, whose result
is well known and constitutes the basis the widely used proximity
force approximation (PFA) \cite{Derjagin}. We use this standard
example to establish the rank among contributions to the total
pressure and show the similarity and differences with the energy
method \cite{opt1}.

We calculate the force acting on the lower plate, denoted by $d$
or \emph{down}, by calculating the pressure on its surface. We
discard the zero and $1d$ (one reflection on the lower plate
itself) reflection terms. The first term to be considered is the
path that bounces once on the upper plate ($u$ or \emph{up}) $1u$.
For parallel plates $\Delta=1/\ell^2$ and we have
\begin{equation}
\label{eq:pressplate}
P(x)= \sum_{r\geq
1u}(-1)^{n_r}\partial_{z}^2\frac{1}{16\pi^2\ell_r^2(x,x)}.
\end{equation}

The length $\ell_r(x,x)$ for the paths that bounce an even number
of times is a constant in $z$ and hence the derivatives vanish:
they do not contribute to the pressure. This seemingly innocuous
observation simplifies the calculations considerably and it is a
test for any other geometry which reduces to parallel plates in
some limit: in this limit the even reflections contributions must
vanish. Generically their contributions are small. This parallels
the role of the odd reflection paths in the energy method
\cite{opt1}.

\begin{figure}
\centerline{\asfigure{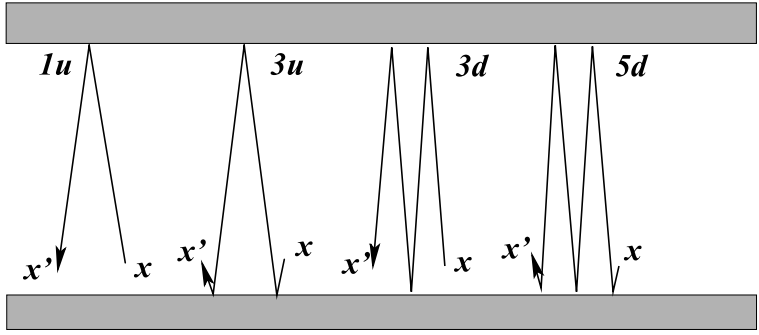}{450}{10cm}}
\caption{\label{fig:paths} Odd reflection paths that contribute to
the Casimir force between the two plates in the pressure
calculations with the optical approximation. The points $x'$ and
$x$ will eventually be taken coincident and lying on the lower
plate.}
\end{figure}

Figure \ref{fig:paths} shows the odd reflection paths labelled
with our conventions.
For the path $1u$ we have
\begin{equation}
\label{eq:plate1u}
P_{1u}(x)=\lim_{z\to
0}-\partial_z^2\frac{1}{16\pi^2(2a-2z)^2}=-\frac{3}{32\pi^2a^4}.
\end{equation}
The next path to be considered is the path that bounces 3 times,
first on $d$, then on $u$ and again on $d$, $dud=3u$ (3 stands for
3 reflections and $u$ for the plate where the middle reflection
occurs) which gives a contribution
\begin{equation}
\label{eq:plate3u}
P_{3u}(x)=\lim_{z\to
0}-\partial_z^2\frac{1}{16\pi^2(2a+2z)^2}=-\frac{3}{32\pi^2a^4}.
\end{equation}
The two contributions Eq.~(\ref{eq:plate1u}) and
Eq.~(\ref{eq:plate3u}) are equal. The reason is easily uncovered.
One can recover Eq.~(\ref{eq:plate3u}) from Eq.~(\ref{eq:plate1u})
sending $z\to-z$ but for the purpose of taking the second
derivative at $z=0$ this is irrelevant. In the same fashion
$P_{3d}=P_{5d}$, $P_{5u}=P_{7u}$ etc.\ and hence we find
\begin{equation}
P(x)=-2\frac{3}{32\pi^2a^4}-2\frac{3}{32\pi^2(2a)^4}-2\frac{3}{32\pi^2(3a)^4}+...=-\frac{3}{16\pi^2a^4}\frac{\pi^4}{90},
\end{equation}
which is the well-known result. Notice also that the rate of
convergence is the same as in the calculation making use of the
Casimir energy in Ref.\ \cite{opt1} ($n$-th term contributes
$1/n^4$ of the first term, in this case $1u+3u$). These
observations that allow us to determine the rank of the
contributions are fundamental, and they apply as well to the other
examples in this section.

\subsection{The Casimir Torsion Pendulum}

In this section we study a geometry already considered in Ref.\
\cite{opt1}: a plate inclined at an angle $\theta$ above another
infinite plate. We have called this configuration a `Casimir
torsion pendulum' because the Casimir force will generate a torque
which can be experimentally measured. The configuration is
analogous to the parallel plates case but the upper plate must be
considered tilted at an angle $\theta$ from the horizontal. The
length of the upper plate must be taken finite, we denote it by
$w$, while the length of the lower plate can be infinite which we
choose for simplicity. There is only one substantial difference
with the parallel plates case: the even reflection paths do
contribute in the pendulum, since their length varies as we move
the final points $x',x$.

We calculate the force exerted on the lower, infinite plate for
simplicity. We then obtain the energy $\cE$, by integrating over
the distance along the normal to the lower plate and from this we
can calculate the torque as
\beq
\cT=-\frac{\p \cE}{\p \theta}.
\eeq

The lower plate is taken infinite, the upper plate width is $w$,
and the distance between the height at the midpoint of the upper
plate is $a$. We will choose as the origin of the coordinates one
point on the intersection line between the lower plate and the
line obtained by prolonging the upper plate. This defines a
fictitious wedge of opening angle $\theta$. We call $x$ the
horizontal and $z$ the vertical coordinate, the third direction,
along which one has translational symmetry, being $y$.

Since the surfaces are locally flat we have $\Delta=1/\ell^2$ as
in the case of the parallel plates, and again the odd reflections
are exactly as in the case of the parallel plates. However now the
even reflections contribute (the notation is the same as in the
parallel plates case, in the even reflections $2u$ means the first
reflection is on the upper plate etc.):
\beq
\label{eq:PserPend}
P=P_{1u+3u}+P_{2u+2d}+P_{3d+5d}+...\
\eeq
where we have grouped the terms with the symbolic notation
$P_{a+b}=P_a+P_b$ when $P_a=P_b$. It is useful to recapitulate
what we have learned about the rank of these contributions:
$P_{1u+3u}$ dominates, $P_{3d+5d}$ is smaller by $\sim 1/16$,
$P_{5u+7u}$ is smaller by $\sim 1/81$, {\it etc.\/}   The even
reflections are generically much smaller than the odd reflections,
and vanish as $\theta\to 0$.

The first term in (\ref{eq:PserPend}) is
\beq
P_{1u+3u}=-2\frac{1}{16\pi^2}\partial_z^2\frac{1}{\ell_1^2(z,x)},
\eeq
with $\ell_1=2(x\sin\theta-z/\cos\theta)$, and an overall factor
of $2$ takes into account the identity $P_{1u}=P_{3u}$. Taking the
derivative and then setting $z=0$ we find
\beq
P_{1u+3u}=-\frac{3}{16\pi^2}\frac{1}{x^4\sin^4\theta\cos^2\theta},
\eeq
and integrating from $x_m=(a/\sin\theta-w/2)/\cos\theta$ to
$x_M=(a/\sin\theta+w/2)/\cos\theta$ we find the force per unit
length in the $y$ direction
\beq
F_{1u+3u}=-\frac{\cos\theta}{32\pi^2\sin^4\theta}\left(\frac{1}{(a/\sin\theta-w/2)^2}-\frac{1}{(a/\sin\theta+w/2)^2}\right).
\eeq
Since term by term $F=-\partial \cE/\partial a$ we find the first
term in optical expansion of the Casimir energy $\cE$ (the
arbitrary constant is chosen so that $\cE\to 0$ when $a\to
\infty$) as
\beq
\cE_{1u+3u}=-\frac{aw{\cos^4 \theta }}
  {2{\pi }^2
    {\left( 4a^2 -
        w^2{\sin^2 \theta }
        \right) }^2}
\eeq
and from this one obtains the torque
\beq
\cT_{1u+3u}= \frac{2aw
    \left(w^2 -4a^2  \right)
    {\cos^3 \theta }
    \sin \theta }{{\pi }^2
    {\left( 4a^2 -
        w^2{\sin^2 \theta }
        \right) }^3}
\eeq
Analogously we can calculate the contribution to the pressure $P$
of the two reflections paths $2u$ and $2d$. Again the
contributions of the two paths are identical and the result
simplifies to
\beq
P_{2u+2d}=\frac{2}{8\pi^2}\frac{1}{2}\partial_z^2\frac{1}{\ell_2^2(z,x)},
\eeq
and using $\ell_2=2\sqrt{x^2+z^2}\sin\theta$ we find
\beq
P_{2u+2d}=-\frac{1}{16\pi^2\sin^2\theta\ x^4}
\eeq
which integrated from $x_m=(a/\sin\theta-w/2)/\cos\theta$ and
$x_M=(a/\sin\theta+w/2)/\cos\theta$ gives the force along the $z$
axis due to these paths:
\beq
F_{2u+2d}=-\frac{\cos^3\theta}{48\pi^2\sin^2\theta}\left(\frac{1}{(a/\sin\theta-w/2)^3}-\frac{1}{(a/\sin\theta+w/2)^3}\right).
\eeq
This expression can now be expanded for $\theta\ll 1$
(quasi-parallel plates)
\beq
F_{2u+2d}\simeq-\frac{1}{16\pi^2}\left(\frac{w}{a^4}\theta^2+\frac{5w^3-11wa^2}{6a^6}\theta^4+...\right).
\eeq
Notice that this expression vanishes when $\theta\to 0$, as it
should since for parallel plates all the contributions of even
reflections paths vanish.

The next term in the series is $F_{3d+5d}$, whose calculation is
performed in the same fashion. The result is:
\begin{eqnarray}
F_{3d+5d}&=&-\frac{3}{16\pi^2}\frac{\cos^5 2\theta}{\sin^4 2
\theta}\left(\frac{1}{(a/\sin\theta-w/2)^3}-\frac{1}{(a/\sin\theta+w/2)^3}\right),\\
&\simeq&-\frac{1}{16\pi^2}\left(\frac{3w}{16a^4}+\frac{5w^3-48a^2w}{32a^6}\theta^2+...\right).
\end{eqnarray}
We can also present the term given by the 4 reflections paths,
\beq
F_{4u+4d}=-\frac{\cos^3 2\theta}{48\pi^2\sin^2
2\theta}\left(\frac{1}{(a/\sin\theta-w/2)^3}-\frac{1}{(a/\sin\theta+w/2)^3}\right)
\simeq-\frac{1}{16\pi^2}\left(\frac{w}{4a^4}\theta^2+...\right).
\eeq
The terms independent of $\theta$ can be seen to reconstruct the
parallel limit case $F=-(1+1/16+1/81+...)3/16\pi^2a^4$.

Term by term, this series for the force reproduces the series in
Ref.~\cite{opt1}. The series for the energy and the torque agree
as well. The results of the pressure method then coincide with
those of the energy method (as for all the examples analyzed in
this paper). In Ref.~\cite{opt1} we discussed at some length the
predictions of the optical method for the Casimir torsion
pendulum. We will not repeat them here, referring the reader to
that paper for further details.

\subsection{Sphere and Plane}
\label{sec:sphereplane}

The sphere facing a plane is an important example for several
reasons: it has been analyzed theoretically with various exact or
approximate numerical techniques \cite{SandS,Gies03}; it is an
experimentally relevant configuration; the exact solution is
unknown and probably will escape analytical methods for a long
time to come. We have already calculated the optical approximation
to the Casimir energy in Ref.~\cite{opt1} up to $5$ reflections.
In this paper we study this problem for mainly pedagogical
purposes, leaving a more accurate and complete numerical analysis
for the future. We believe it is worth studying this example
because, contrary to the previous two examples, the enlargement
factor plays an important role and moreover we will reanalyze this
example with finite temperature in Section IV B 2.

We calculate the pressure (and by integrating, the force) exerted
on the plate by the sphere which, of course, equals the force
exerted by the plate on the sphere. We start from the qualitative
observation that the rank of the contributions is the same as in
the parallel plates case in the limit $a/R\to 0$. In all the
examples we have analyzed this rank is preserved for any value of
$a/R$. Moreover the ratios of the contributions to the force
$F_{3+5}(a,R)/F_{1+3}(a,R)$, $F_{4}(a,R)/F_{2}(a,R)$ etc.\
decrease quickly as $a/R$ increases, we believe due to the growing
importance of the enlargement factor.

In this paper we calculate analytically the $1s$ term (here $s$
stands for `sphere' and $p$ for `plate') and by using the relation
$P_{1s+3s}\equiv P_{1s}+P_{3s}=2P_{1s}$ proved in Section
\ref{sec:enmomtens} (the notation is the same as in that section)
we are able to include the $3s$ term as well.

Using the expressions for the length and enlargement factor for
the $1s$ path obtained in Ref.~\cite{opt1} we get
\barr
P_{1s+3s}&=&-2\frac{R}{16\pi^2}\frac{\partial^2}{\partial
z^2}\frac{\Delta^{1/2}_{1s}}{\ell_{1s}}\nonumber\\
&=&-\frac{R}{32\pi^2}\frac{\partial^2}{\partial z^2}
    \left( R - {\sqrt{{\left( a + R -
              z \right) }^2 + {\rho }^2}}
      \right)^{-2}\left({\left( a + R -
              z \right) }^2 + {\rho }^2\right)^{-1/2}\bigg|_{z=0}.
\label{eq:P13sphere}
\earr
The final expression for the pressure $P_{1s+3s}$ obtained after
the derivatives are taken is rather long, however the contribution
to the force on the plate, $F_{1s+3s}$ (obtained by integration of
$P_{1s+3s}$ over the infinite plate) is quite simple:
\beq
\label{eq:F1s3s}
F_{1s+3s}=2\pi\int_0^\infty d\rho \rho P_{1s+3s}=-\frac{\hbar c
R}{8\pi a^3}.
\eeq
This is the largest of the contributions and increasing $a/R$
improves the convergence of the series due to the presence of the
enlargement factor, so the asymptotic behavior at large $a/R$
predicted by the optical approximation is that given by this
formula, \emph{i.e.}\ $F\propto R/a^3$ or $E\propto R/a^2$. This
asymptotic law is in accordance with the numerics of
Ref.~\cite{pap1} and the predictions of other semiclassical
methods \cite{SandS}. However, eq.~(\ref{eq:F1s3s}) is in
disagreement with the Casimir-Polder law \cite{CasimirPolder}
which predicts $E\propto R^3/a^4$ for $a\gg R$. This is no great
surprise, since our method is not valid for $a/R\gg 1$, the
semiclassical reflections being corrected and eventually
overshadowed by diffractive contributions
\cite{Keller,SandSdiffr}.

We have calculated the contribution of the two reflections paths
analytically as well. The calculation is more involved than the
one reflection term but a big simplification occurs if one notices
that, for the purpose of taking the second derivative with respect
to $z$ at $z=0$, one can leave the reflection point on the sphere
fixed. We could not prove a similar result for any other
reflection. It is certainly not true for \emph{odd} reflections
but one can conjecture it to be true for \emph{even} reflections.
In this paper we have not calculated the 4 reflection terms and
hence we could not check this conjecture for more than 2
reflections.

And finally, we have calculated the $3p$ (or $sps$) and hence
obtained the $5p$, or $pspsp$, paths contribution $P_{3p+5p}$;
$P_{3p+5p}$ in the parallel plates limit should account for $\sim
1/16$ of the total force. This contribution, unlike the previous
ones, must be calculated partly numerically, mainly because
finding the reflection point on the sphere requires the (unique)
solution of a transcendental equation. This task is achieved much
more quickly by a numerical algorithm than by patching together
the several branches of the analytic solution.
\begin{figure}
\centerline{\hspace{-3cm}\asfigure{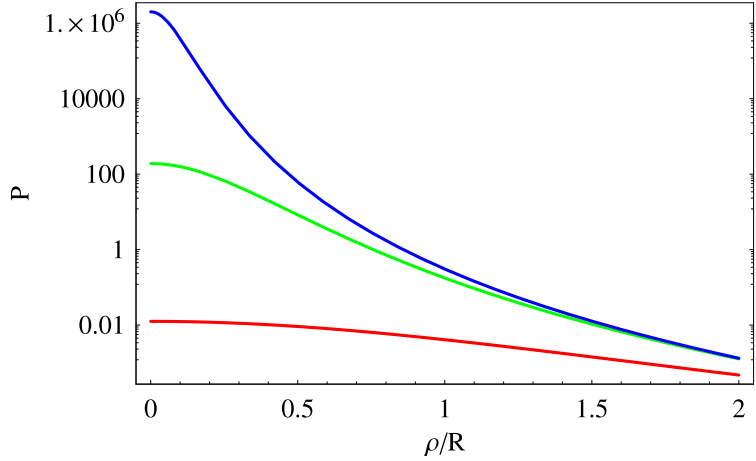}{550}{12cm}}
\caption{\label{fig:pressure-sp} The magnitude of the total
pressure up to reflection $5p$ in units of $\hbar c/R^4$ as a
function of the radial coordinate on the plate, $\rho/R$. Upward,
or red to blue $a/R=1$, $a/R=0.1$ and $a/R=0.01$. }
\end{figure}
\begin{figure}
\centerline{\hspace{-3cm}\asfigure{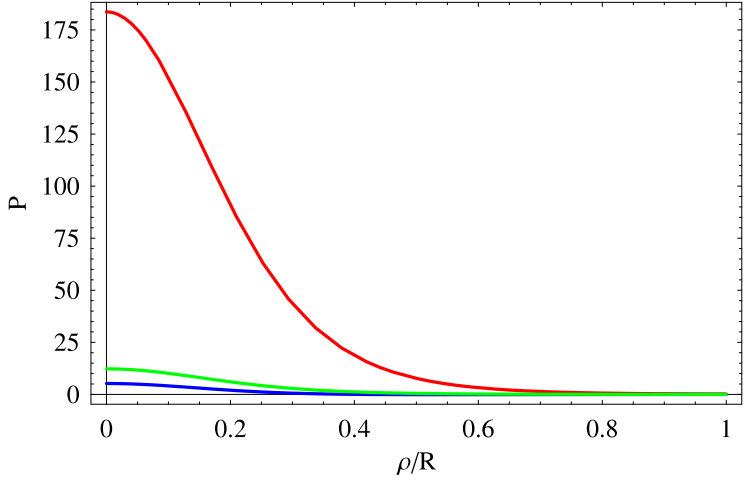}{550}{12cm}}
\caption{\label{fig:presspic} Contributions to the pressure in
units of $\hbar c/R^4$ as a function of $\rho/R$, for fixed
$a/R=0.1$. Downward or red to blue, we have $-P_{1s+3s}$,
$-P_{3p+5p}$ and $P_{2+2}$. Although unnoticeable in this figure,
the curve $P_{2+2}$ changes sign at around $\rho/R\simeq 0.4$ (see
Figure \ref{fig:P2negative} for a similar situation).}
\end{figure}
\begin{figure}
\centerline{\hspace{-3cm}
\asfigure{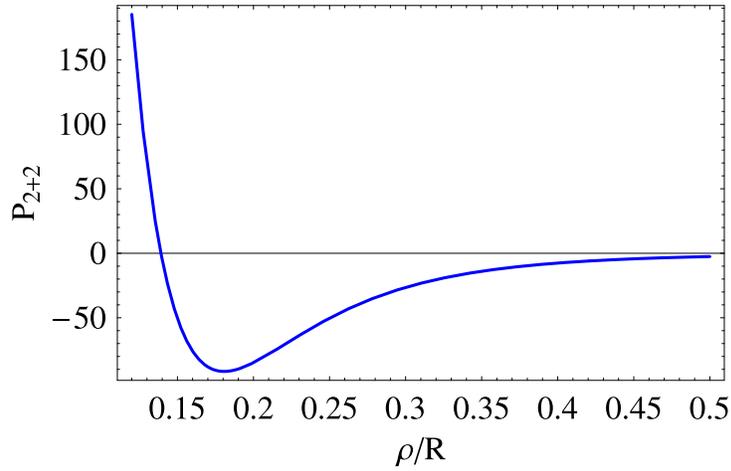}{550}{12cm}}
\caption{\label{fig:P2negative} Contribution of the two reflection
path(s) to the pressure in units of $\hbar c/R^4$ as a function of
$\rho/R$, for fixed $a/R=0.01$. The pressure becomes negative,
showing that the sign of the pressure is not determined by the
number of reflection only.}
\end{figure}

The total pressure is plotted in Fig.~in \ref{fig:pressure-sp}
while the various contributions (keeping in mind that $P_{1s+3s}$
and $P_{3p+5p}$ are negative and $P_{2+2}$ is mainly positive) are
shown in Fig.~\ref{fig:presspic}. Figure \ref{fig:pressure-sp}
reveals some interesting features of the pressure in this
geometry: the total pressure decays very quickly with the distance
as $P\sim \rho^{-\alpha}$: the exponent $\alpha$ seems to depend
upon the distance $a/R$, but for $a/R\leq 0.1$ a good fit is
obtained with $\alpha=6$, in accordance with the asymptotic
expansion of the $1+3$ reflection term Eq.\ (\ref{eq:P13sphere});
by decreasing the distance between the sphere and the plate, the
pressure becomes more and more concentrated near the tip, giving
us reasons to trust our approximation and supporting the use of
the PFA as a first approximation in the limit $a/R\to 0$.
Figure~\ref{fig:presspic} shows the relative importance of the
contributions due to the different paths. As expected the
contribution to the total pressure decreases quite fast by
increasing the number of reflections. In Fig.~\ref{fig:P2negative}
one can also see that the sign of the pressure is not determined
simply by the number of reflections of the underlying optical path
--- as for the contribution to the energy density.
\begin{figure}
\centerline{\hspace{-3cm}\asfigure{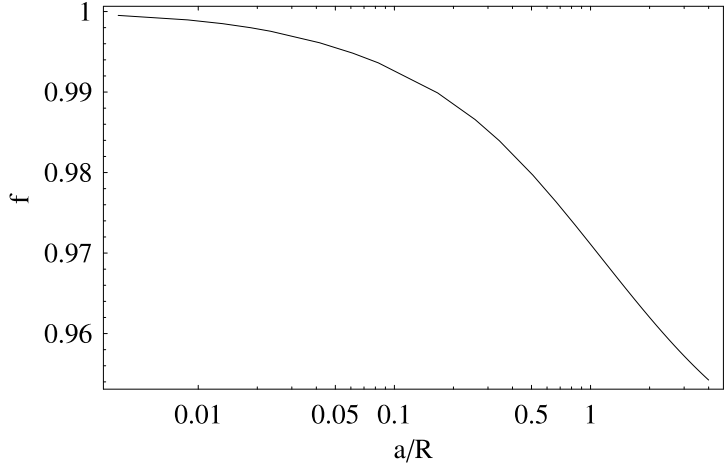}{550}{12cm}}
\caption{\label{fig:f} The ratio between the optical force up to
the $5p$ reflection and the most divergent term in the PFA, as
defined by eq.~(\ref{def}).}
\end{figure}

By integrating the pressure over the whole plate we obtain the
force $F$. It is useful to factor out the most divergent term of
the force, as predicted by the PFA, so we define the quantity
$f(a/R)$ as
\beq
\label{def}
F(a)=-\frac{\pi^3 R}{720a^3}f(a/R).
\eeq
Since we include only a finite number of reflections it is
convenient to factor out the constant $\zeta(4)/(1+1/16)$ such
that $f$ is normalized with $f(0)=1$. The function $f(a/R)$,
calculated including paths $1s,\ 3s,\ 2,\ 3p$ and $5p$, is plotted
in Figure \ref{fig:f}. When $a/R\to 0$ $f$ is fitted by
\beq
\label{eq:fsphere}
f(a/R)=1-0.10\ a/R+\Ord{(a/R)^2}\ .
\eeq
By comparing to the results of \cite{pap1}
\beq
\label{eq:fsphereen}
f_{\rm energy}(a/R)=1+0.05\ a/R+\Ord{(a/R)^2}\,
\eeq
there is the difference in the sub-leading term.

By neglecting the $5s+7p$ reflection paths (which in the parallel
plates case contribute $\sim 2\%$ of the total force) we can only
assert that the functions $f$ in (\ref{eq:fsphere}) and
(\ref{eq:fsphereen}) represent the optical approximation with an
error of $2\%$. When plotted on the whole range of $a/R$ where the
optical approximation is to be trusted the pressure and energy
method curves never differ more than $2\%$. However there is no
such a bound on the sub-leading term which, on the contary,
depends on the higher reflections contributions which have not
been included in this calculations.\footnote{For example consider
that including only $1s,~3p$, and 2 and reflections would have
given a sub-leading term $-0.16a/R$ instead of $-0.10a/R$ in
Eq.~(\ref{eq:fsphere}). The sub-leading term then changes of
$50\%$ by adding the $3s+5p$ reflection terms which contributes
only up to $8\%$ of the total.} With the terms calculated at this
point, we cannot make a precise statement about the sub-leading
term. We can however safely say that the subleading term $a/R$
coefficient is quite small and our method disagrees with the PFA
prediction $-0.5a/R$. The sphere opposite plate is such an
experimentally relevant geometry that further, more accurate
studies need to be performed to compare with experimental data.

In conclusion, the lessons to be learned from this example are
two: 1) The calculations with the pressure method are even quicker
and simpler than the energy method and sometimes can give analytic
results for non-trivial geometries and 2) the sub-leading terms
must be compared only between calculations performed with the same
accuracy.\footnote{AS would like to thank M.\ Schaden and S.\
Fulling for conversations on this point during the workshop
`Semiclassical Approximations to Vacuum Energy' held at Texas A \&
M, College Station, TX, January 2005. The concerns about the
errors to be associated with the optical, semiclassical or
proximity force approximation is still open to debate and is
strictly connected to one of the most challenging open problems in
spectral theory \emph{i.e.\ }how to go beyond the semiclassical
approximation to the density of states of a positive Hermitian
operator.}


\section{Casimir Thermodynamics.}
\setcounter{equation}{0} \label{sec:thermo}

As measurements of Casimir forces increase in accuracy they become
sensitive to thermal effects.  The natural scale for Casimir
thermodynamics is a distance, $\tilde\beta=\hbar c/\pi T$, which
at room temperature is about 2.5 microns.  [To avoid confusion
with the wave number $k$, we set Boltzmann's constant equal to
unity and measure temperature in units of energy.  We continue to
keep $\hbar$ and $c$ explicit.]  So, assuming the corrections are
of $\Ord{(a/\tilde\beta)^\alpha}$, depending on the value of
$\alpha$ thermal effects might be expected between the $10\%$ (for
$\alpha=1$) and $0.3\%$ (for $\alpha=4$, the standard parallel
plates result) level for Casimir force measurements on the micron
scale. In open geometries, like the sphere and plane, even longer
distance scales are probed by Casimir effects, and this gives rise
to interesting changes in the temperature dependence of the
Casimir free energy in comparison with the case of parallel
plates\cite{MT}.  The optical approximation is well suited for
discussion of thermodynamics since the thermodynamic observables,
like the Casimir energy, can be expressed in terms of the
propagator.  Here we consider again a non-interacting, scalar
field outside rigid bodies on which it obeys Dirichlet boundary
conditions.

Before entering into a technical discussion of temperature
effects, it is useful to anticipate one of our central results
which follows from qualitative observations alone.  As $T\to 0$
the temperature effects probe ever longer distances. Even at room
temperature the natural thermal scale is an order of magnitude
larger than the separation between the surfaces in present
experiments (see Ref.~\cite{expt1}).  Since long paths contribute
little to the Casimir force, we can be confident that thermal
effects vanish quickly at low temperature. However, the leading
$T$-dependence at small $T$ comes from regions beyond the range of
validity of the optical (or any other) approximation, so we are
unable to say definitively how they vanish for geometries where no
exact solution is possible ({\it i.e.\/} other than infinite
parallel plates).

This section is organized as follows:  First we discuss the free
energy and check our methods on the parallel plates geometry; then
we discuss the temperature dependence of the pressure, which we
apply to the sphere and plate case.  Finally we discuss the
difficulties associated with the $T\to 0$ limit.

\subsection{Free Energy}

The free energy is all one needs to calculate both thermodynamic
corrections to the Casimir force and Casimir contributions to
thermodynamic properties like the specific heat and pressure.
However like the Casimir \emph{energy}, Casimir contributions to
the specific heat, pressure, {\it etc.\/}, are cutoff dependent
and cannot be defined (or measured) independent of the materials
which make up the full system.  So we confine ourselves here to
the thermal corrections to the Casimir force. The problem of
parallel plates has been addressed before and our results agree
with those\cite{MT}.

\subsubsection{Derivation}

We start from the expression of the free energy for the scalar
field as a sum over modes
\begin{eqnarray}
\cF_{\rm
tot}&=&-\beta^{-1}\sum_n\ln\left(\frac{e^{-\beta\frac{1}{2}\hbar\omega_{n}}}
{1-e^{-\beta(\omega_{n}-\mu)}}\right),\nonumber\\
&=&\beta^{-1}\sum_n\ln\left(1-e^{-\beta(\hbar\omega_{n}-
\mu)}\right)+\sum_{n}\frac{1}{2}\hbar\omega_n,\nonumber\\
&\equiv&\cF+\cE,
\end{eqnarray}
where $\mu$ is the chemical potential, and the last term is the
Casimir energy, or the free energy at zero temperature, since
$\cF=0$ for $T=0$. The Casimir energy $\cE$, being independent of
the temperature, does not contribute to the thermodynamic
properties of the system. It however does contribute to the
pressures and forces between two bodies. The force between two
bodies, say $a$ and $b$, is obtained by taking the gradient of the
free energy with respect to the relative distance $\vecr_{ab}$
\begin{equation}
    \vf_{ab}=-\vnabla_{ab} \cF.
\end{equation}
At $T=0$ we recover the familiar result $\vf=-\vnabla \cE$.

Next we turn the sum over modes into a sum over optical paths.
Following the same steps that led from Eq.~(\ref{eq:T00}) to
Eq.~(\ref{eq:T00opt}) we obtain
\beq
    \cF=\beta^{-1}\int d^Nx\ \int_0^\infty dk\
    \rho(x,k)\ln\left(1-e^{-\beta (\hbar\omega(k)-\mu)}\right).
\eeq
where $\rho(x,k)$ is given by Eq.\ (\ref{eq:rhoxk}). By
specializing to a massless field in $3$ dimensions with zero
chemical potential (to mimic the photon field), and substituting
the optical approximation for the propagator Eq.~(\ref{eq:Gopt3}),
we obtain the sum over paths
\beq
\label{eq:Fropt}
    \cF\equiv\sum_{r=0}\cF_{r}
    =\sum_{r}(-1)^{r}\frac{1}{2\pi^{2}\beta}\int_{\cD_r}d^3x\ \Delta_r^{1/2}
    \int_0^\infty
    dk \ k \sin(k\ell_r) \ln\left(1-e^{-\beta\hbar c
    k}\right).
\eeq
Here the term $\cF_{0}$, the direct path, gives the usual free
energy for scalar black body radiation.  Using the values for the
direct path, we have $\Delta_0=1/\ell_0^2$   and
$\ell_{0}=|x'-x|\to 0$ when taking $x'\to x$. We get the familiar
textbook expression
\begin{equation}
\cF_{0}=V\int_0^\infty dk\ \frac{k^2}{2\pi^2}\
\beta^{-1}\ln\left(1-e^{-\beta\hbar
ck}\right)=-\frac{\pi^2}{90}\frac{VT^4}{(\hbar c)^3},
\label{eq:F0classic}
\end{equation}
where $V$ is the (possibly infinite) volume outside the bodies.

The general term $\cF_{r}$ associated with the path $r$ is
calculated by performing the $k$ integral in Eq.~(\ref{eq:Fropt}):
\begin{equation}
    \label{eq:Fr}
    \cF_r=(-1)^{r+1}\frac{\hbar c}{2\pi^{2}}\int_{\cD_r}d^3x
    \Delta_r^{1/2}\frac{1}{2\ell_r^3} \left[-2+ \tilde\ell_r
    \left({\rm coth} \tilde\ell_r +\tilde\ell_r {\rm
    csch}^2\tilde\ell_r \right)\right]
\end{equation}
where $\tilde\ell_{r}=\ell_{r}\pi T/\hbar c =
\ell_{r}/\tilde\beta$ measures the path length relative to the
thermal length scale.

Eq.~(\ref{eq:Fr}) is the fundamental result of this section and
gives a simple, approximate description of thermal Casimir effects
for geometries where diffraction is not too important. There are
no divergences in any of the $\cF_{r}$, ultraviolet or otherwise,
even for the direct path (as we saw in eq.\ (\ref{eq:F0classic}))
and the first reflection path. All the ultraviolet divergences are
contained in the Casimir energy $\cE$. Indeed, by expanding the
integrand of equation (\ref{eq:Fr}) at short distances,
\emph{i.e.} $\tilde\ell_{r}\ll 1$, we obtain
\begin{equation}
    \label{eq:frsmallt}
     \Delta_r^{1/2}
     \frac{1}{2\ell_r^3}\left[-2+\tilde\ell_r \left({\rm
      coth} \tilde\ell_r   +\tilde\ell_r {\rm
      csch}^{2}\tilde\ell_r  \right)\right]
    \simeq
     \Delta_r^{1/2}\frac{1}{\tilde{\beta}^3}
     \left[\frac{1}{45\tilde{\beta}}
     \ell_r-\frac{4}{945\tilde{\beta}^3}\ell_r^3+...\right].
\end{equation}
Only the 1-reflection path length can go to zero to generate a
divergence. For this contribution $\Delta_{r}$ diverges like
$1/\ell_{r}^2$ as $\ell_{r}\to 0$, however this is compensated
 by  the $\ell_r$ term in (\ref{eq:frsmallt}) so the expression is
finite and then integrable.

To check for infrared divergences notice that at large distances,
$\tilde\ell_{r}\gg 1$, the integrand of (\ref{eq:Fr}) goes to
$\sim \Delta_{r}^{1/2}/\ell_{r}^{2}$. For an infinite flat plate
the $\Delta_{r}\sim 1/z^{2}$, where $z$ is the normal coordinate
to the plate, and the integral is hence $\sim dz/z^{3}$ at large
$z$. For finite plates the domain of integration is finite and for
curved plates the enlargement factor falls even faster than
$1/\ell^2$, and the integral remains convergent.

Since the integral converges in both the infrared and ultraviolet,
it is safe to estimate the important regions of integration by
naive dimensional analysis.  This leads to the conclusion that
\emph{The paths that dominate the temperature dependence of the
Casimir force have lengths of order the thermal length
$\tilde\beta$}.  High temperature implies short paths.  Very low
temperatures are sensitive to very long paths.  Long paths involve
both paths experiencing many reflections, which are sensitive to
the actual dynamics at and inside the metallic surface, or paths
making long excursions in an open geometry, which are sensitive to
diffraction.  Either way, low temperatures will present a
challenge.

\subsubsection{Parallel Plates}

We know that in the limit of infinite, parallel plates the optical
approximation to the propagator becomes exact. Hence our method
gives another way  to calculate the free energy of this
configuration of conductors. It is convenient to study this
example to check against known results and to prepare the way for
a study of the $T\to 0$ limit.

We recall that for this configuration the expression for the
enlargement factor is $\Delta=1/\ell^2$ and the lengths are given
by $\ell_{2n}=2na$ (where $a$ is the distance between the plates)
and $\ell_{2n+1,\ u}=2(a-z)+2na$, $\ell_{2n+1,\ d}=2z+2na$, the
notation being the same as in Section \ref{sec:parplates}, should
at this point be familiar to the reader.

As in the zero temperature case it is useful consider even and odd
reflection contributions separately and as for the zero
temperature case, the sum over odd reflections   turns into an
integral over $z$ from $0$ to $\infty$
\begin{equation}
    \cF_{{\rm odd}}=\sum_{n=0}^\infty \cF_{2n+1, d}+\cF_{2n+1, u}=\frac{\hbar c}{2\pi^2\tilde{\beta}^3}S\int_0^\infty
    dx \frac{1}{2x^4}\left[-2+x({\rm
      coth} x +x\ {\rm csch}^2 x )\right],
\end{equation}
where $x=2z/\tilde{\beta}$ and $S$ is the area of the plate. The
definite integral can be easily performed numerically and its
value is $\nu=0.06089...$,
\begin{equation}
    \label{eq:foddpp}
    \cF_{{\rm odd}}= 2\frac{\hbar
    c}{4\pi^2\tilde{\beta}^3}S\nu=\frac{\pi T^3}{2(\hbar c)^2}S\nu
\end{equation}
which is independent of the separation, $a$, and therefore does
not contribute to the force.

Let us turn now to the even reflection paths. They have constant
length $2na$, so the volume integral simply yields the volume
between the surfaces $v=Sa$. We already calculated the
zero-reflection term $\cF_0$ in Eq.~(\ref{eq:F0classic}). The
remaining even reflection contributions (2,4,6,...\ reflections)
$\cF_{{\rm even}, r\geq 2}$ can be written as an infinite sum
\begin{equation}
\label{eq:Freven}
    \cF_{{\rm even}, r\geq 2}=-2\frac{\hbar c}{2\pi^2}Sa\frac{1}{\tilde{\beta}^4}
    \sum_{n=1}^\infty\frac{1}{2x_n^4}\left[-2+x_n\left({\rm coth}
    x_n+x_n{\rm csch}^2 x_n\right)\right]
\end{equation}
where $x_n=2na/\tilde{\beta}\equiv n\tau$ (this defines the
dimensionless temperature $\tau$) and we have introduced an
overall factor of two to take into account the multiplicity of the
paths.  Thus the total free energy for parallel plates is the sum
of $\cF_{0}$ (eq.~(\ref{eq:F0classic}) and the results of
eqs.~(\ref{eq:foddpp}) and (\ref{eq:Freven})),
\begin{equation}
\label{eq:pptot}
    \cF_{\parallel} =-\frac{\pi^4}{90}\frac{VT^4}{(\hbar c)^3}  +\frac{\pi T^3}{2(\hbar c)^2}S\nu - \frac{\hbar c}{ \pi^2\tilde {\beta}^4}Sa \sum_{n=1}^\infty
 \frac{1}{2x_n^4}\left[-2+x_n\left({\rm coth}
    x_n+x_n{\rm csch}^2 x_n\right)\right]
\end{equation}

It is not possible to rewrite $\cF_{\parallel}$ in a closed form,
but the sum is easy to compute numerically and the high and low
temperature expansions are easy to obtain analytically. At high
temperatures (and fixed $a$) $\tau\to\infty$, and the summand
$g(n)$ in eq.~(\ref{eq:pptot}) falls rapidly enough with $n$
\begin{equation}
g(n)=\frac{1}{2(\tau n)^4}\left[-2+(\tau n)\left({\rm coth}
    (\tau n)+(\tau n){\rm csch}^2 (\tau
    n)\right)\right]=\frac{1}{2(\tau n)^4}\left[-2+\tau
    n\right]+\Ord{e^{-\tau n}},
\end{equation}
that the limit may be taken under the summation, with the result,
\begin{equation}
    \cF_{{\rm even},r\geq 2}\simeq-\frac{\hbar
    c}{\pi^2\tilde{\beta}^4}Sa\sum_{n=1}^\infty\left[
    \frac{1}{2n^3\tau^3}-\frac{1}{\tau^4n^4}\right] =-\frac{\zeta(3)}{16\pi
    a^2}ST+ \frac{\pi^2\hbar c}{1440a^3}S.
\end{equation}
Notice that the second term cancels the even paths contribution to
the Casimir energy. Hence the final expression for the high $T$
expansion of the free energy is particularly simple,
\begin{equation}
\label{eq:cFtothight}
    \cF_{\rm tot}= \cF + \cE = -\frac{\pi^{2}}{90\hbar c}VT^{4}+\nu\frac{\pi}{2(\hbar c)^{2}}ST^{3}
    -\frac{\zeta(3)}{16\pi a^{2}}ST+\Ord{e^{-\pi T a/\hbar
    c}}.
\end{equation}
The first term is usual black body contribution to the bulk free
energy. It does not contribute to the force.  The second term is
also independent of $a$ and does not give rise to any force. The
third term instead gives the thermal Casimir force. Notice that
$\hbar c$ has disappeared from this expression.  Called the
``classical limit'', this high temperature behavior has been noted
before and some early results are even due to Einstein (in
\cite{Milonni} pg.\ 2; see also \cite{Feinberg}). In the next
section, after the thermal corrections to the pressure are
calculated, we show how to extend this result to other geometries.

Note some interesting features of the $T\to\infty$ limit:  First,
the sum over paths converges like the sum of $(1/n)^{3}$ as
indicated by the appearance of $\zeta(3)$.  While slower than the
$T=0$ convergence, it is still rapid enough to obtain a good
approximation from low reflections.  Second, note that the
$T\to\infty$ problem in 3-dimensions corresponds exactly to a
$T=0$ problem in 2-dimensions. This is an example of the familiar
dimensional reduction expected as $T\to\infty$. We can give a
short proof of this result. Let us first write:
\beq
F=-\frac{1}{\beta}\log Z
\eeq
where $Z$ is the partition function. We need to evaluate $Z$ to
the lowest order in $\beta$ when $\beta\to 0$. The thermal scalar
field theory can be written as a free theory on the cylinder
$\mathbb{R}^3\times [0,\beta)$. For $\beta\to 0$ the dynamics
along the thermal coordinate is frozen in the ground state, with
energy $E_0=0$, where $\phi$ does not depend on the thermal
coordinate. The partition function $Z$ is now
$Z=Z_{3}+\Ord{e^{-\beta E_1}}$ where $E_1$ is the first excited
state $E_1\propto 1/\beta^2$ and $Z_{3}$ is the partition function
of the remaining three-dimensional problem in $\mathbb{R}^3$. If
the conductors geometry is symmetric along one spatial coordinate,
say $x$ (in the parallel plates problem we have two of these
directions, $x$ and $y$) this can now be interpreted as an
Euclideanized time variable extending from $0$ to $L_x/c$. So we
will write $Z_3=Z_{2+1}=e^{-\frac{1}{\hbar}\cE_2L_x/c}$ where
$\cE_2$ is the Casimir energy of the 2 dimensional problem of two
lines of length $L_y$, distant $a$. The free energy $F$ is then:
\beq
\label{eq:F21}
F=-\frac{1}{\beta}\log Z\simeq-\frac{1}{\beta}\log
Z_{2+1}=T\frac{1}{\hbar}\frac{L_x}{c}\cE_{2}=-TL_xL_y\frac{\zeta(3)}{16\pi^2a^2}.
\eeq
Since $S=L_xL_y$ This is exactly the $a$-dependent term in
eq.~(\ref{eq:cFtothight}). If the geometry is not translational
invariant then we can only say from eq.~(\ref{eq:F21}) that the
free energy is linear in $T$ (since $Z_{2+1}$ is independent of
$\beta$). Later, by using the optical approximation we will find
an explicit analytic expression valid also for non-symmetric,
smooth geometries.

For low temperatures, $\tau\to 0$,  the terms in the $n$-sum in
eq.~(\ref{eq:pptot}) differ very little from each other so we can
use the Euler-McLaurin formula\cite{AS},
\begin{equation}
    \sum_{n=1}^\infty g(n)=\int_0^\infty dx\
    g(x)-\frac{1}{2}g(0)-\frac{1}{12}g'(0)+... \ =
    \frac{\nu}{\tau}-\frac{1}{90}+\Ord{\tau} .
    \label{eq:em}
\end{equation}
Substituting into eq.~(\ref{eq:pptot}) we find that the first term
in eq.~(\ref{eq:em}) cancels the sum over odd reflections (the
second term in eq.~(\ref{eq:pptot})) and that the second term in
eq.~(\ref{eq:em}) combines with $\cF_{0}$ to give a very simple
result,
\begin{equation}
\label{eq:excludeV}
\cF_{\rm tot}=\cE-\frac{(V-Sa)\pi^2T^4}{90(\hbar c)^3}.
\end{equation}
at low temperatures.  This has a simple physical interpretation:
the typical thermal excitations of the field at low temperature
have very long wavelengths, it is hence energetically inconvenient
for them to live between the two plates. As a result the only
modification of the $T=0$ result is to exclude from the standard
black body free energy the contribution from the volume between
the plates.    One could imagine measuring this effect as a
diminished heat capacity for a stack of conducting plates inside a
cavity.

The low temperature result, eq.~(\ref{eq:excludeV}), is
deceptively simple.  Its simplicity obscures an underlying problem
with the $T\to 0$ limit.  We postpone further discussion until we
have explored the temperature dependence of the pressure.  Suffice
it to say for the moment, that eq.~(\ref{eq:excludeV}) probably
does not apply to realistic conductor with finite absorption,
surface roughness, and other non-ideal characteristics.

    \subsection{Temperature dependence of the pressure}
\label{sec:PT}

In this section we will obtain the temperature dependence of the
pressure within our approximation and apply it to  a preliminary
study of the sphere and plate case. To begin, we calculate the
thermal average of an operator $\cO$ quadratic in the real scalar
field $\phi$. The average of a generic operator $\cO$ is given by
the trace over a complete set of eigenstates $\ket{\Psi_\alpha}$
of the Hamiltonian weighted by a Boltzmann factor:
\beq
\label{eq:tavg}
\tavg{\cO}=\sum_\alpha e^{-\beta
\cE_\alpha}\bra{\Psi_\alpha}\cO\ket{\Psi_\alpha}.
\eeq
After some algebra we find
\barr
\tavg{\cO}&=&\sum_{j}\cO_{j}\tavg{2n_{j}+1}\nonumber\\
&=&\sum_{j}\cO_{j}\frac{1+e^{-\beta E_j}}{1-e^{-\beta E_j}}
\earr
where $\tavg{}$ denotes the thermal average, $j$ labels the normal
modes $\psi_j$ (cf.~Section II), $n_j$ is the occupation number of
the mode $j$ and $E_j$ its energy. The quantities $\cO_j$ are read
from the decomposition of the diagonal part of the operator $\cO$
written as $\cO_{\rm diag}=\sum_j \cO_j(a_j^\dag a_j+a_j
a^\dag_j)$ where $a_j$ is the annihilation operator of the mode
$j$.

The $\cO_j$ for the pressure can be read easily from the analysis
in Section \ref{sec:enmomtens}:
\beq
P_j=\lim_{x'\to
x\in\cS}\frac{1}{4E_{j}}\partial'_{\vn}\partial_{\vn}\psi_j(x')\psi_j(x)
\eeq
So we can write the pressure on the plate at non-zero temperature
as
\begin{eqnarray}
\label{eq:temppress}
P(x\in\cS)&=&\lim_{x'\to
x}\sum_j\frac{1}{4E_{j}}\partial'_n\partial_n\psi_j(x')\psi_j(x)\left(\frac{1+e^{-\beta
E_{j}}}{1-e^{-\beta E_{j}}}\right)\nonumber\\
&=& \lim_{x'\to x}\partial'_{\vn}\partial_{\vn}\int_0^\infty dk
e^{-k/\Lambda} \frac{k}{2\pi E(k)}\Im\
G(x',x,k)\left(\frac{1+e^{-\beta
E(k)}}{1-e^{-\beta E(k)}}\right)\nonumber \\
&=&\Im \int_0^\infty dk e^{-k/\Lambda}\frac{k}{2\pi
E(k)}\frac{1}{2}\partial^2_{\vn}\ G(x,x,k)\left(\frac{1+e^{-\beta
E(k)}}{1-e^{-\beta E(k)}}\right)
\end{eqnarray}
where we have used Eq.~(\ref{eq:trick1}).

Next we introduce the optical approximation for the propagator and
limit ourselves to massless scalars $E(k)=\hbar c k$. The
discussion of the divergences parallels that of Section
\ref{sec:enmomtens} and needs not be repeated here. We remove
$P_0$ and $P_1$ and leave all the finite contributions $r\in\cR$.
The optical approximation for the pressure exerted by a massless
scalar field reads
\begin{eqnarray}
\label{eq:Ptemp}
P(x)&=&\sum_{r\in\cR}(-1)^{n_r}\partial^2_{\vn}\frac{\Delta_r^{1/2}}{16\pi^2}\left[\frac{1}{\tilde{\beta}}
\coth\left(\ell_r/\tilde{\beta}\right)\right].
\end{eqnarray}
where it is understood that the zeroth and first reflection terms,
which contribute to the pressure on each surface individually, but
not to the force between surfaces, have been dropped.

Before applying this to the sphere and plate problem, let us again
look at the limiting behavior as $T\to\infty$ and $T\to 0$, and
draw some conclusions independent of the detailed geometry.  First
consider $T\to\infty$.  The shortest paths in the sum in
eq.~(\ref{eq:Ptemp}) are of order $a$, the intersurface
separation.  [Remember that the optical approximation is accurate
as long as the important paths are short compared to $R$, a
typical radius of curvature of the surfaces.]  At high $T$ we can
take the $\tilde\beta\to 0$ limit under the sum over reflections
since the resulting sum still converges.  Therefore low
reflections dominate, and we can see, retrospectively, that the
high temperature approximation applies when $\tilde\beta/a\to 0$.
So as $T\to\infty$,
\beq
\label{eq:highT}
P=\sum_r(-1)^{n_r}\partial^2_{\vn}\frac{\Delta_r^{1/2}}{16\pi^2}
\left[\frac{1}{\tilde{\beta}}+\Ord{\frac{1}{\tilde{\beta}}e^{-\ell_r/\tilde{\beta}}}\right].
\eeq
This limit has been called (it has been previously found for the
parallel plates case) the ``classical limit''
\cite{Milonni,Feinberg,MT}, since the final expression for high
temperatures, reinserting $\hbar$ and $c$,
\beq
\label{eq:classicallimit}
P\simeq\sum_r(-1)^{n_r}\partial^2_{\vn}\frac{\Delta_r^{1/2}}{16\pi}T
\eeq
is independent of $\hbar$ and $c$ apart from exponentially small
terms. This expression amounts in neglecting the 1 in the
expression $\tavg{2n_j+1}$, corresponding to normal ordering or
neglecting the contribution of the vacuum state.

At low temperatures, $\tilde\beta\to\infty$, it is not possible to
interchange the limit with the sum.  The relevant quantity is
$\frac{1}{\tilde\beta}\coth(\ell_{r}/\tilde\beta)$, which goes
like
\beq
\label{eq:lowT}
\frac{1}{\tilde\beta}\coth\left(\frac{\ell_{r}}{\tilde\beta}\right)=
\frac{1}{\ell_r} + \frac{\ell_r}{3\,{\tilde{\beta} }^2} -
  \frac{\ell_r^3}{45\,{\tilde{\beta} }^4} +
  \Ord{\frac{\ell_r^5}{\tilde{\beta}^4}}
\eeq
as $\tilde\beta\to\infty$.  The first term yields the familiar
$T=0$ expression.  The others would give divergent contributions
because of the factors of $\ell_{r}$ in the numerators (even after
the inclusion of the enlargement factor $\Delta_r$). Of course the
sum over reflections of the \emph{difference},
$\frac{1}{\tilde\beta}
\coth(\ell_{r}/\tilde\beta)-\frac{1}{\ell_{r}}$, converges to zero
as $\tilde\beta\to\infty$, so thermal corrections definitely
vanish for any geometry as $T\to 0$ as expected.  Once again we
relegate more detailed consideration of the $T\to 0$ limit to a
later subsection.

\subsubsection{Sphere and plate}
\label{sec:sphplateT}
In this section we calculate the pressure and total force for the
configuration of a sphere facing a plane at non-zero temperature
within $5p$ reflections.   The optical approximation should be
accurate if the important paths are short compared to $R$, the
radius of the sphere.  On the other hand the thermal corrections
to the force are sensitive to paths with lengths of order
$\tilde\beta$.  So we must have $R\gg \tilde\beta$ and $R \gg a$
in order to obtain reliable results from the optical
approximation.  Fortunately this is a region of experimental
interest:  present experiments use, for example, $a\approx 0.5\mu
m$, $R\approx 100\mu m$, and at room temperature,
$\tilde\beta\approx 2.5\mu m$.  In this regime the optical
approximation should give a good description of the thermal
corrections to the force between perfectly reflective, perfectly
smooth conductors.
\begin{figure}
\centerline{\asfigure{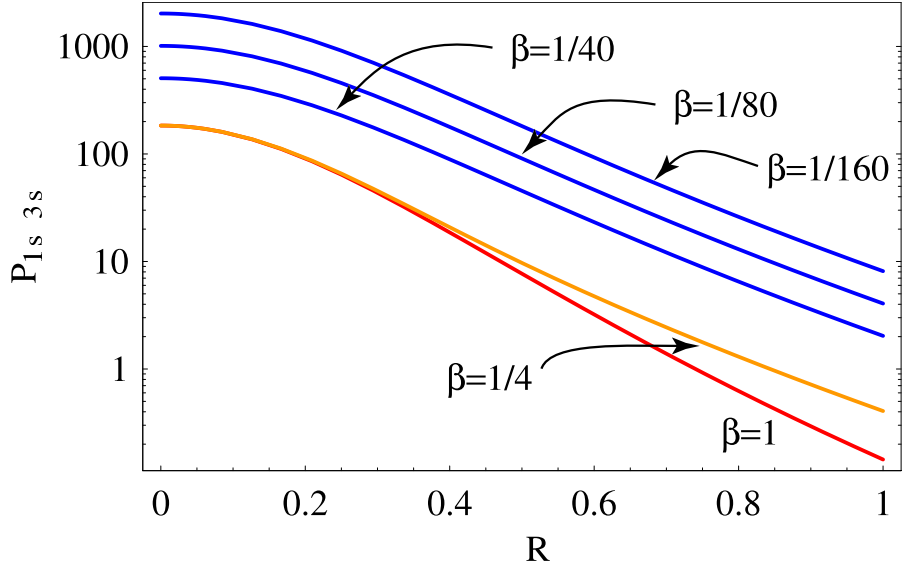}{650}{12cm}}
\caption{\label{fig:p13temp} The $\rho$ dependence of the $1s+3s$
contribution to the pressure $P_{1s+3s}$ for the sphere and the
plate in units of $\hbar c/R^4$ for various temperatures. Two
effects must be noticed. The top 3 curves (in blue) show the
high-temperature region where the pressure is proportional to $T$
(notice the logarithmic scale). The two lower curves (in orange
and red) show the low-temperature region when increasing the
temperature changes the asymptotic behavior of $P$ for large
$\rho$ (\emph{i.e.\ }$\rho\gtrsim \tilde\beta$) while for small
$\rho$ the behavior reduces to the zero-temperature limit.}
\end{figure}

The expression for the pressure is given by Eq.~(\ref{eq:Ptemp}),
the enlargement factors and lengths are the same as in the $T=0$
case. By applying Eq.~(\ref{eq:Ptemp}) to the $1s+3s$ paths we
find the results in Figure \ref{fig:p13temp}. Notice that at high
temperatures increasing the temperature essentially scales the
whole plot proportionally to $T$. The force is then linearly
dependent on the temperature (this is the `classical limit'
already discussed in Section \ref{sec:PT}). More details are given
in the caption of Figure \ref{fig:p13temp}.

\begin{figure}
\centerline{\hspace{-3cm}\asfigure{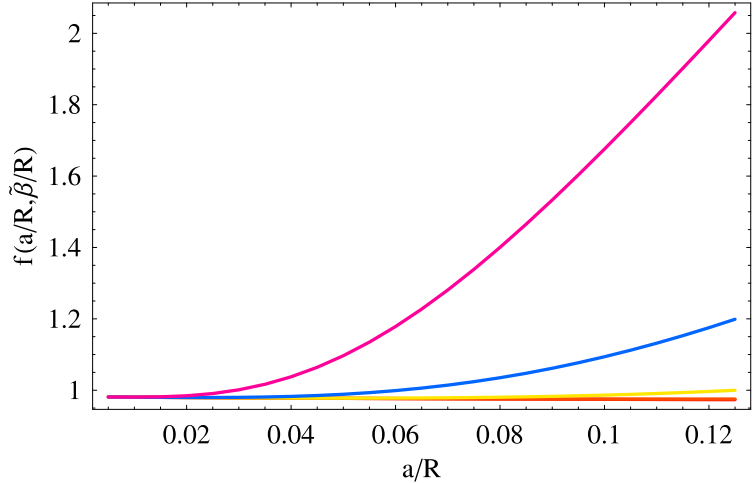}{650}{12cm}}
\caption{\label{fig:fT} The function $f(a/R,\tilde{\beta}/R)$ as a
function of $a/R$ for $\tilde{\beta}/R$ (from red to violet or
down up) $=1,1/2,1/4,1/8,1/16$. $f(0)\simeq 0.98$ since we are
summing only up to reflection $5p$. The two lowermost curves, red
and orange ($\tilde{\beta}=1,1/2$) superpose almost exactly.}
\end{figure}

A dimensionless function $f(a/R,\tilde{\beta}/R)$ can again be
defined by rescaling the total force $F$ to extract the leading
term as $a\to 0$. The limiting behavior $a\to 0$ is not affected
by temperature effects so we stick to the old definition for $f$:
\beq
\label{def:ftherm}
F(a,\tilde{\beta},R)=-\frac{\hbar c \pi^3 R}{720
a^3}f(a/R,\tilde{\beta}/R).
\eeq
In Figure \ref{fig:fT} we present $f$ (up to 5 reflections) for 5
different values of $\tilde{\beta}/R$ (we choose 1, 1/2, 1/4, 1/8
and 1/16 recognizing that $\tilde\beta\sim 1$ strains the limits
of our approximations) and varying $a$. Notice that in a
neighborhood of $a/R=0$, shrinking as $\tilde\beta/R$ increases,
the function $f$ is very well approximated by the $T=0$ form,
already discussed in Section \ref{sec:sphereplane}, $f(a/R)\simeq
1-0.1 a/R$. It is not useful to study the derivative
$A(\tilde{\beta}/R)=\partial f(x,\tilde{\beta})/\partial x$ as
$x=a/R\to 0$ since this will take the constant value predicted by
the zero temperature analysis, or $-0.1$ in this approximation,
for any value of the temperature we choose.

It is also clear from the previous discussions leading to equation
(\ref{eq:highT}) that in the opposite regime, for
$a/\tilde{\beta}\gg 1$, we must have $F\propto R / a^2
\tilde{\beta}=R T/a^2$ (the `classical limit'). In fact, the first
term in the high temperature expansion (\ref{eq:highT}) integrated
over $\rho$ converges and gives a finite force linear in $T$. For
this problem, the first term in the reflection expansion for high
temperatures can even be calculated analytically:
\begin{equation}
F_{1s+3s}=-\hbar c\frac{R}{8\pi
a^2\tilde{\beta}}+\Ord{e^{-R/\tilde\beta}}\simeq-\hbar c\frac{R}{8
a^2} T.
\end{equation}
Unfortunately there is no such simple closed expression for higher
reflection terms (nor for this first term at arbitrary $T$).
However, if one believes that the rank of contributions is similar
to the parallel plates case one should feel safe to say that this
truncation captures the optical approximation within a
$\zeta(3)-1\simeq 20\%$. Hence our statements are at least
\emph{qualitatively} correct.

This expression for the force gives a prediction for the function
$f$, defined in Eq.\ (\ref{def:ftherm}). At this level of accuracy
($1s+3s$ reflection) and for $a/\tilde{\beta}\gtrsim 1$, apart for
exponentially small terms in the temperature expansion we have
\beq
\label{eq:highTf13}
f_{1s+3s}\simeq\frac{90}{\pi^4}\frac{a}{R}\frac{R}{\tilde{\beta}}
\eeq
which grows linearly in $a/R$ and is (interestingly enough)
independent of $R$. This is evident in Fig.\ \ref{fig:fT} for the
curves with $\tilde{\beta}=1/8,\ 1/16$. For higher $\tilde{\beta}$
the linear growth starts at higher values of $a$ not shown in
Fig.\ \ref{fig:fT}. Moreover the exponential accuracy manifests
itself in the sudden change of behavior from $f\simeq 1-0.1a/R$ to
$f\propto a/\tilde{\beta}$.

It is quite easy to extract a universal prediction from this data,
whatever the definitive numbers are, after the sum over optical
paths is carried to sufficiently high order: \emph{for any
non-zero temperature the function $f(a/R)$ will deviate from his
zero-temperature behavior at $a\gtrsim\tilde{\beta}\sim\hbar c/T$.
The deviation will be in the upward direction, increasing the
attractive force between the bodies.} Eventually, for sufficiently
large distances, the high temperature behavior given by eq.\
(\ref{eq:classicallimit}) (or (\ref{eq:highTf13}) for the
sphere-plane problem) will be recovered.

\subsection{Thermal corrections at low temperatures}

The preceding examples have made it clear that in the language of
the optical approximation, thermal corrections at low temperature
arise from very long paths, $\ell_{r}\sim \tilde\beta$.  This can
be seen from the general form of the free energy,
eq.~(\ref{eq:Fr}), or in the attempt to take the
$\tilde\beta\to\infty$ limit under the summation in
eq.~(\ref{eq:Ptemp}), which fails because of the expansion,
eq.~(\ref{eq:lowT}).  Here we examine this non-uniformity more
carefully in general and in particular for the parallel plate
case, where all the expressions are available.  We then attempt to
draw some conclusions about the magnitude of corrections at low
temperature and the possibility of calculating them reliably in an
model that idealizes the behavior of materials.

We return to eq.~(\ref{eq:temppress}), which gives the exact
expression for the pressure, and separate out the thermal
contribution,
\beq
\label{eq:dPtot}
P(T)-P(0)\equiv\delta P=\Im\int_0^\infty
dk\frac{1}{2\pi}\partial^2_{n'n}\cG(x',x,k)2\frac{e^{-\beta \hbar
c k}}{1-e^{-\beta \hbar c k}},
\eeq
still exact. Expanding the denominator in a geometric series, we
find
\beq
\delta P = \frac{1}{\pi}\Im\sum_{m=1}^{\infty}\int_0^\infty dk
\partial^2_{n'n}\cG(x',x,k) e^{-m\beta \hbar c k}.
\eeq
Each term in the sum is a Laplace transform of the Greens
function. Clearly, as $\beta\to\infty$ the frequencies that
dominate this integral are $\propto 1/\beta\sim T$.

What are the low frequency contributions to $\cG(x',x,k)$?  In the
ideal case of infinite, perfectly conducting, parallel plates,
there is a gap in the spectrum at low $k$: $k\ge\frac{\pi}{a}$.
However \emph{in realistic situations} the plates are finite
and/or curved, the geometry is open, and there is no gap in the
spectrum.  The low-$k$ part of the spectrum is sensitive to the
global geometry, including edges and curvature, and to the  low
frequency properties of the material.  If the conditions are close
to the ideal, the contributions to $\delta P$ from small $k$ may
be small.  However as $T\to 0$, they dominate.  We conclude that
the $T\to 0$ behavior of $\delta P$ cannot be calculated for
realistic situations.

The optical approximation does not take account of diffraction,
and cannot accurately describe the $T\to 0$ limit.  Nevertheless
it is interesting to see how it fails, since this sheds light on
the problem in general. Substituting the optical expansion for the
Greens function (replacing
$\partial^2_{n'n}\to\frac{1}{2}\partial^2_z$ and setting
$\hbar=c=1$) we find
\barr
\delta P&=&-\sum_{m=1}^{\infty}\sum_{r\geq
1}\frac{1}{8\pi^2}\partial^2_z\int_0^\infty dk
\Delta_r^{1/2}\sin(k\ell_r)e^{-m\beta k}\nonumber\\
&=&-\sum_{m=1}^{\infty}\sum_{r\geq
1}\frac{1}{8\pi^2}\partial^2_z\Delta_r^{1/2}\frac{\ell_r}{m^{2}\beta^2+\ell_r^2}.
\earr
The problems with $T\to 0$ are quite apparent:  as
$\beta\to\infty$ all paths become important.

Next we specialize to parallel plates where $\ell_r=(2ar\pm 2z)$.
The derivative can be carried out explicitly.  For simplicity we
focus on $m=1$  ($\delta P =\sum_{m=1}^{\infty}\delta P_{m}$),
\beq
\delta P_{1}=-
\frac{2}{\pi^2}\sum_{r=1}^{\infty}\frac{12(ar)^2-\beta^2}{\left(4(ar)^2+\beta^2\right)^3},
\eeq
which can be rewritten using the variable $\tau=2\pi a/\beta$
introduced earlier,
\beq
\label{sumform}
\delta P_{1}=- \frac{2\pi^{2}}{
\beta^{4}}\sum_{r=1}^{\infty}\frac{3\tau^2r^2-\pi^{2}}{(
\tau^2r^2+\pi^{2})^3}.
\eeq
The sum can be performed, giving
\beq
\delta
P_{1}=-\frac{1}{\pi^2}\Bigg(\frac{1}{\beta^4}-\frac{\pi^3}{8a^3\beta}\coth\left(\frac{\pi\beta}{2a}\right)
{\rm csch}^2\left(\frac{\pi\beta}{2a}\right)\Bigg).
\eeq
The second term in brackets is exponentially small as
$\beta\to\infty$.  If we ignore it, restore the $m$-dependence,
and sum over $m$, we obtain
\beq
\label{eq:dPfin}
\delta P=-\frac{\pi^2}{90\beta^4}.
\eeq
which agrees with our earlier calculation, as it must.

However eq.~(\ref{sumform}) allows us to study the convergence of
the sum over reflections as $\beta\to\infty$.  Instead of
performing the sum analytically, we sum up to some $r_{\rm
max}\equiv X$.  Since $\tau\to 0$, we can once again use
Euler-Maclaurin,
to rewrite the sum over $r$ as
\beq
\label{partsum}
\delta
P_{1}=-\frac{2}{\pi^2}\frac{1}{\beta^4}\left[\frac{1}{2}-\frac{X}{(1+\tau^2X^2/\pi^{2})^2}+
\frac{1}{2}\frac{3\tau^2X^2/\pi^{2}-1}{(1+\tau^2X^2/\pi^{2})^3}+
...\right].
\eeq
where the omitted terms are higher Euler-Maclaurin contributions
that are unimportant as $\beta\to\infty$ (\emph{i.e.\ }$\tau\to
0$).

If the upper limit on the sum, $X$, is taken to $\infty$, only the
first term, $1/2$, survives and gives the expected result.  The
question is:  How large must $X$ be before the limiting behavior
set in?  Dropping the third term in eq.~(\ref{partsum}), which is
subdominant, we can rewrite $\delta P_{1}$ as
\beq
\delta P_{1}
=-\frac{2}{\pi^2\beta^4}\left[\frac{1}{2}-\frac{X}{(1+
X^2\tau^2/\pi^{2})^2}\right]=-\frac{2}{\pi^2\beta^4}\left[\frac{1}{2}+\frac{1}{\tau}f(\tau
X)\right].
\eeq
The function $f(z)$ is negative definite and has a minimum at
$z=1/2\sqrt{3}\simeq 0.29$ where it takes the value
$-3^{3/2}/32\simeq -0.16$. So in order the result
Eq.~(\ref{eq:dPfin}) to be valid we must include $X\gg
X_c=\pi/\sqrt{3}\tau$ terms in the sum. For example in a typical
experimental situation we have $a=0.5\mu m$ and $T=300 K$ so
$\beta=8\mu m$, $\tau=8/\pi$ and $X_c=8/\sqrt{3}=4.6$. In this
case it is necessary to go to $X\sim 20$ before the contribution
of $|f(\tau X)/\tau|$ is smaller than 1/2.  This means paths with
$\sim 40$ reflections and path lengths of order $20\mu m$. With 40
chances to sample the surface dynamics of the material and paths
of $20\mu m$ available to wander away from the parallel plate
regime, the idealizations behind the standard parallel plates
calculation must be called into question.

It must be said however that in the modern experiments the
temperature corrections are at most of the order of a few percent
at $a\sim 1\mu m$ and vanish when $a\to 0$. Nonetheless we want to
point out that there is a conceptual difference between
formulations based on the infinite parallel plates approximation,
extended to curved geometries by means of the PFA, and a
derivation (like ours) in which the curvature is inserted \emph{ab
initio}. The thermal and curvature scales interplay in a way that
the usual derivations \cite{MT,MPnew} could not possibly capture,
giving rise to different power law corrections in $a/\beta$. It is
worth reminding the reader that the usual numerical estimates of
thermal corrections are based on the infinite parallel plates
power law $(a/\beta)^4$. A smaller power like $(a/\beta)^2$ would
give a much bigger upper bound.

To summarize:  temperature corrections are small at small $T$, but
the existing methods of calculating them, including both our
optical approximation and the traditional parallel plates
idealization, cannot be trusted to give a reliable estimate of the
$T$-dependence at small $T$.

\section{Conclusions}

In this paper we have shown how to adapt the optical approximation
to the study of local observables.  We have illustrated the method
by studying the pressure, but the method applies as well to other
components of  the stress tensor, to charge densities, or any
quantity that can be written in terms of the single particle
Greens function. The advantage of the optical approximation is to
extend the study of these local observables to novel geometries.
 In particular we developed an expression for the Casimir pressure
on the bodies and applied our main result Eq.\ (\ref{eq:press2})
to the study of three important examples: parallel plates, the
Casimir pendulum and a sphere opposite a plate.

We have also shown how to calculate, within this approximation
scheme, thermodynamic quantities and thermal corrections to the
pressure in the general case and applied our results to the
example of parallel plates (retrieving the known results) and to
the case of a sphere opposite a plate.  Along the way we have
given a proof of the ``classical limit'' of Casimir force for any
geometry (within our approximation), \emph{i.e.\ }the fact that
Casimir forces at high temperatures are proportional to the
temperature and independent of $\hbar$, a fact that previously was
known only for parallel plates.

Finally, we argued that all known methods of computing the
temperature dependence of the Casimir effect are suspect as $T\to
0$.

\section{Acknowledgments}

We would like to thank S.~Fulling and M.~Schaden for comments. AS
would like to thank M.~V.~Berry for useful discussions. RLJ would
like to thank the Rockefeller Foundation for a residency at the
Bellagio Study and Conference Center on Lake Como, Italy, where
much of this work was performed.  This work is also supported in
part by funds provided by the U.S.~Department of Energy (D.O.E.)
under cooperative research agreement DE-FC02-94ER40818.


\end{document}